\documentclass[aps,prx,twocolumn,groupedaddress,nofootinbib]{revtex4-2}

\usepackage{hyperref}
\usepackage{CJK}
\usepackage{amsmath}
\usepackage{graphicx}
\usepackage{dcolumn}
\usepackage{enumerate,amsthm,amssymb,color}
\usepackage{bbm}
\usepackage{bm}
\usepackage[normalem]{ulem}
\usepackage[french, english]{babel}
\usepackage{synttree}
\usepackage{caption}
\usepackage{subfig}
\usepackage{multirow}
\usepackage{xfrac}
\usepackage{fixltx2e}
\usepackage{comment}

\newcommand{\beq}{\begin{equation}}
\newcommand{\eeq}{\end{equation}}
\newcommand{\bea}{\begin{align}}
\newcommand{\eea}{\end{align}}
\newcommand{\bq}{\begin{quote}}
\newcommand{\eq}{\end{quote}}
\newcommand{\rob}{\color{black}}
\newcommand{\robs}{\color{black}}

\newcommand{\blk}{\color{black}}

\begin{document}
\title{Reply to ``Comment on `Experimentally adjudicating between different causal accounts of Bell-inequality violations via statistical model selection'\! ''}

\author{Patrick Daley}

\author{Kevin J. Resch}
\affiliation{Institute for Quantum Computing and Department of Physics \& Astronomy, University of Waterloo, Waterloo, Ontario N2L 3G1, Canada}

\author{Robert W. Spekkens}\email{rspekkens@perimeterinstitute.ca}
\affiliation{Perimeter Institute for Theoretical Physics, 31 Caroline Street North, Waterloo, Ontario Canada N2L 2Y5}


\begin{abstract}
Our article described an experiment that adjudicates between different causal accounts of Bell inequality violations by a comparison of their predictive power, finding that certain types of models that are structurally radical but parametrically conservative---of which a class of superdeterministic models are an example---overfit the data relative to models that are structurally conservative but parametrically radical in the sense of endorsing an intrinsically quantum generalization of the framework of causal modelling. In their comment (arXiv:2206.10619), Hance and Hossenfelder argue that we have misrepresented  the purpose of superdeterministic models.  We here dispute this claim by recalling the different classes of superdeterministic models we defined in our article and our conclusions regarding which of these are disfavoured by our experimental results.
Their confusion on this point seems to have arisen in part from the fact that we characterized superdeterministic models within a causal modelling framework and from the fact that we referred to this framework as ``classical'' in order to contrast it with an intrinsically quantum alternative.  In this reply, therefore, we take the opportunity to clarify these points. They also claim that if one is adjudicating between a pair of models, where one model can account for strictly more operational statistics than the other, the first model will tend to overfit the data relative to the second.  Because this model inclusion relation can arise for pairs of models in a reductionist heirarchy, they conclude that overfitting should not be taken as evidence against the first model. We point out here that, contrary to this claim, one does {\em not} expect overfitting to arise generically in cases of model inclusion, so that it is indeed sometimes appropriate to consider overfitting as a criterion for adjudicating between such models.
\end{abstract}

\maketitle

\section{Introduction}


The main criticism of our article~\cite{daley22} in the comment~\cite{hance24} by Hance and Hossenfelder (hereafter HH)  is that we ``potentially misrepresent the very purpose of the superdeterministic models present in the literature.''  HH provide two reasons in defence of this assessment:
(i) they claim that we ``[selected] classical superdeterministic models as the ones most worth analysing'', 
 and that this class of superdeterministic models excludes the ones of interest in the literature; (ii)
they claim that our finding that certain superdeterministic models overfit the data relative to the class of models that endorse  an intrinsically quantum common cause (which we termed the \textsc{qCC} class in our article)
 does not imply that one should disfavour the former relative to the latter because
 ``overfitting, while better as a measure of finetuning than other measures given in the literature, does not necessarily indicate a model is universally bad.''


In our reply, we will dispute each of these points. 
Along the way,
we respond to some other criticisms of HH and make further clarifying remarks.

\section{Confusion regarding the scope of supdeterministic models we consider and our usage of the term ``classical''}


HH assert:
``Daley et al state that they analyze classical superdeterministic models using their method,'' and 
 ``Superdeterministic models are in general not classical in any meaningful sense of the word,'' and
 ``the purpose of superdeterminism is not to return to classical mechanics.'' They add: ``It is somewhat unclear why Daley et al would focus their analysis on models with this property [returning to classical mechanics], given that it is ostensibly not fulfilled in most models presented in the literature.''

What HH mean by the phrase ``return to classical mechanics'' is unclear.  
Presumably they mean a return to the {\em operational predictions} of classical physics, in which case they are claiming that we inappropriately confined our attention to the class of superdeterministic models that were {\em operationally classical}.
In support of this interpretation is the fact that when HH contrast  the superdeterministic models presented in the literature with the ``classical'' class that they impute to us, they highlight the feature of the superdeterministic models that is relevant for computing operational predictions, namely, their use of the standard formalism of operational quantum theory (``wave-functions and density matrices, superpositions and entanglement''), suggesting that they take the class we endorse to be one that is
{\em inconsistent} with computing predictions from this formalism. \blk

However, this claim---that the class of superdeterministic models we consider fails to reproduce the predictions of operational quantum theory---is mistaken. 
   In our article, we stated explicitly that we were considering only causal models that {\em do} reproduce all such predictions,
    and we even took the trouble to distinguish two ways in which this can occur: (i) the causal model could realize  all  {\em and only} the data tables predicted by operational quantum theory, a case which in our appendix we termed a {\em quantum-on-the-nose} causal model (which must necessarily involve a restriction on the parameters of the model in the conventional framework for causal modelling), and (ii) the causal model could realize all of the data tables predicted by operational quantum theory {\em together with others that are not}, a case which in our appendix we termed a  {\em quantum-extending} model.  \rob A quantum-extending model might or might not involve a restriction on the parameters of the model in the conventional framework for causal modelling; we introduced the terms {\em parameter-restricted} and {\em parameter-unrestricted} to distinguish these two cases. \blk
      The quantum-extending class includes  precisely the sort of superdeterministic models that HH favour, as we will argue below.


 HH's mistake concerning our view  seems to have arisen in part from a misconception regarding 
 our usage of the term ``classical'' in certain discussions.   We begin, therefore, by correcting this misconception.
 \blk



First,  note that different researchers have different opinions about what principles, concepts, and frameworks within classical physics must be modified in quantum physics.
  Some researchers, for example, have explored the idea that the innovation of quantum physics relative to classical physics might be best understood as an innovation to {\em logic}.  
 One such proposal is  that in order to accommodate quantum physics, the partial order of propositions under the ordering relation of logical implication is not a Boolean lattice but an orthomodular lattice, wherein the distributive law
   fails but a weaker law termed ``orthomodularity'' still holds~\cite{hooker2012logico}.
  Researchers working on this program naturally came to
   use the term ``classical logic'' to refer to conventional logic and ``quantum logic'' to  refer to the exotic alternative being proposed to accommodate quantum physics.
    Consequently, when describing an interpretation of quantum theory that  {\em does not} endorse a modification of logic (which is most of them), it is appropriate to say
      that it is classical {\em in its logic}, i.e.,  that it is {\em logically} classical.


Similarly, our article was framed within the context of a research program that explores the idea that what must be modified by virtue of physics being quantum rather than classical  are the principles and concepts of {\em causal modelling}.
Rather than representing causal relata by sets and causal relations by functions, as is conventionally done,
 one contemplates an exotic alternative, such as representing these by operator algebras and unitaries respectively~\cite{schmid2020unscrambling}.
And rather than representing inference (i.e., what one can infer about one causal relata given another) by
 conditional probability distributions, one contemplates doing so with an exotic alternative, such as conditional density operators \cite{leifer13}.  It is then natural for researchers pursuing this research program to use the term ``classical framework for causal modelling'' to refer to the conventional framework and ``quantum framework for causal modelling'' to refer to the exotic alternative to it.
 This is precisely the sort of classical-quantum dichotomy we were discussing in our article.

In short, stipulating that a given research program assumes the classicality of some particular concept, principle, or framework---such as classical logic or the classical framework for causal modelling---is not to stiuplate that this program implies a `return to classical mechanics'.  Rather, it is only to clarify, for that research program, which concepts, principles, and frameworks are {\em not} being revised in the face of quantum physics, in order to better highlight which {\em are}.
   


Thus, contrary to HH's claim that superdeterministic models are ``not classical in any meaningful sense of the word'', one can identify many parts of the theoretical framework assumed by superdeterminists where it is the conventional alternative that is endorsed rather than the exotic one.
 Superdeterministic models endorse classical logic  for instance.  They also endorse the classical framework for causal modelling.   The latter point is what we emphasized in our article.

Specifically, we noted that in causal modelling one can choose to be radical (i.e., embrace something exotic) on one of two fronts: (i) the causal structure, or (ii) the parameters that one adds to this structure.\footnote{In principle, one could be radical on {\em both} fronts, but one front is sufficient and all existing proposals avail themselves of the possibility of remaining conservative on one of them.}   Regarding different hypotheses about the causal structure underlying the Bell experiment, we stipulated that to endorse the one where the only causal connection between the two wings is a common cause acting on the outcomes (the causal structure assumed by Bell) is to be conservative about the causal structure, while allowing a cause-effect influence between the wings, 
or positing that the hidden variable might influence one or both of the setting variables \rob  (which is how one describes the superdeterminist's hypothesis in the causal modelling framework) \blk
  is to be radical about the causal structure.
Meanwhile, to be conservative about the parameters is to endorse the conventional framework for causal modelling, i.e., a ``classical causal model'', whereas to be radical is to endorse an exotic alternative, i.e., a ``quantum causal model''.
As we note in our article, because
a  superdeterministic model is radical about the causal structure, it can be conservative about the nature of the parameters, i.e., it can \rob (and does) \blk endorse the conventional framework for causal modelling.
As such, it is an example of a classical causal model.\footnote{One might even say that for researchers unwilling to countenance the possibility of superluminal causal influences, it is the commitment to the classical framework for causal modelling (implicit or explicit) that leads them to entertain such a radical supposition as superdeterminism.} 
The terminology we introduced sought to capture these facts: ``We refer to such models as {\em parametrically classical} and {\em structurally superdeterministic}, abbreviated as \textsc{cSD}.''
HH's phrase ``classical superdeterministic model'', despite also fitting with the acronym $\textsc{cSD}$, \robs appears nowhere in our article and is misleading insofar as it obscures the fact that we are picking out a {\em particular aspect} of the framework as classical.\blk

In Sec. II of their article. HH articulate what they take to be the essence of a superdeterministic model as follows: it is a hidden variable model that preserves Bell's notion of locality while abandoning the notion of measurement independence (the statistical independence of the pair of setting variables and the hidden variable) and that reproduces Born's rule in certain limits by averaging over the ensemble of possible values of the hidden variables.
  But this is nothing more than an example of a causal model in the \textsc{cSD} class described above. 
This follows from two facts that were first highlighted in Ref.~\cite{wood15}: (i) that a model in the conventional (i.e., classical) framework for causal modelling is ultimately just another way of speaking about a hidden variable model, with parameters of the causal model capturing the distributions associated to each hidden variable, and (ii) that the satisfaction of Bell's notion of locality and the possibility of violating measurement independence can be {\em derived} as a consequence of the causal structure \rob being of the type we called structurally superdeterminist \blk
 (the structure assumed by Bell supplemented with a causal influence from the hidden variable to one or both of the setting variables).

As an aside, we note that in our view it is {\em more apt} to characterize the assumption of superdeterminism in terms of causal structure than in terms of a violation of measurement independence. This is because the causal structure provides an {\em explanation} of the violation of measurement independence: there is a correlation between the hidden variable and a setting variable {\em because} there is a causal influence of the hidden variable on the setting variable. 
Achieving causal explanations of correlations is a critical part of the superdeterminist research program because its central motivation is salvaging Bell's notion of local causality,
and what would be the point of  this salvage operation if one didn't care about explaining correlations causally? 


Allowing models in the \textsc{cSD} class to be  {\em quantum-extending} means allowing parameter values that can lead in principle to deviations from the predictions of quantum theory.  HH favour models that allow for such deviations (outside certain limits wherein the Born rule is reproduced), and so their view is well-characterized by the quantum-extending class of \textsc{cSD} models.  Because the details of HH's preferred superdeterminstic model are not completely worked out \rob (they state ``we don't know exactly under which circumstances the new physics appears''), \blk it is unclear whether it is meant to be parameter-restricted or parameter-unrestricted.

Our article did the data analysis for parameter-unrestricted versions of these models and showed that these overfit the data relative to models in the \textsc{qCC} class (Bell's causal structure with an intrinsically quantum framework for causal modelling). 
As we noted in the article, our data analysis technique can also be applied to the case of a parameter-restricted model and the verdict about overfitting in principle could change since a parameter-restricted model has fewer opportunities for overfitting relative to its parameter-unrestricted counterpart. 
Consequently, anyone with a concrete such model who is able to articulate its content as a restriction on the parameters of a \textsc{cSD} model could apply our methods to see whether or not 
 their model also overfits the data relative to the \textsc{qCC} model.

It is in this sense that our results leave open the possibility that a parameter-restricted quantum-extending superdeterminstic model could outperform the \textsc{qCC} model relative to our train-and-test methodology.  It remains only for proponents of superdeterminism to stipulate the nature of the parameter restriction that one should assume in the analysis. This possibility was made explicit in our article.  

In summary, we can find no basis for HH's claim that we have misrepresented the purpose of superdeterminist models.

\blk





\blk

\section{More expressive power does not necessarily imply overfitting}

We turn now to considering HH's claims regarding why our experiment found overfitting for the superdeterminist model that we considered and why, in their view,
such overfitting
``does not necessarily indicate a model is universally bad.''



HH imply that one should expect to find that a model $M$ overfits the data relative to a model $M'$ whenever $M$ and $M'$ are both part of a reductionist heirarchy, with $M$ 
\rob proving a more fine-grained description \blk
 than $M'$,  as in the case where $M$ is statistical mechanics and $M'$ is thermodynamics.
After describing \rob Richmann's law of mixtures, an equation that gives the final temperature that results when two bodies with different initial temperatures are brought into contact and allowed to equilibrate (their Eq. (1)), \blk they say of it:
\begin{quote}
If all you want to do is fit [the equation],
 then having multiple free parameters for each single atom in the fluid is clearly unnecessary. If you have that many parameters, they will fit every single fluctuation, however minor --- you can make it fit everything. [...] it's overfitting [...]
\end{quote}
They then draw out what lesson they think this holds for our analysis:
\begin{quote}
[...] overfitting for a given scenario is not necessarily bad in all situations, and given superdeterministic models are often used to try to consider physics beyond quantum mechanics, we should expect them to overfit a test such as the one Daley et al perform.
\end{quote}

%

While it is true that positing microscopic degrees of freedom that underlie thermodynamic degrees of freedom is {\em not required} to provide a fit to data describing a few of the thermodynamic variables, it is not the case  that positing such microscopic degrees of freedom necessarily leads to overfitting.
More precisely, If $M$ and $M'$ are two models that can fit the data, and $M$ has strictly more expressive power than $M'$---so that $M$ and $M'$ stand in a relationship of {\em model inclusion}---
 this {\em does not imply} that $M$ will necessarily overfit the data relative to $M'$ in the sense of having a lower training error and a higher test error in a train-and-test methodology.  Overfitting generally occurs 
as a result of a model {\em mistaking statistical fluctuations for real features}, and having more expressive power does not guarantee that such mistakes will be made.


 An example of this phenomenon (model inclusion without overfitting) is actually provided in Appendix C.3 of our article,  \robs  where we describe a dephased version of our experiment wherein the \textsc{qCC} model does not overfit the data relative to the \textsc{cCC} model (the model that is structurally common-cause and parametrically classical) even though they stand in a relationship of model inclusion.  
In Appendix~\ref{app}, we describe this example in detail and make some further observations about the relationship between model inclusion and overfitting.  

\blk



\section{Experimental requirements for seeing deviations from operational quantum theory}

\rob We also disagree with HH \blk regarding the experimental requirements for seeing deviations from operational quantum theory.   As noted above, HH seem to endorse a quantum-extending model.  Conventional experiments, such as the one reported in our article, take place in a limit where deviations from the predictions of operational quantum theory \rob are expected by HH to \blk become small in their model.
They assert that, \rob as a consequence, \blk in order to see a deviation from Born's rule in a superdeterministic theory ``you have to look at a system which is outside that limit.''
 Being outside this limit, however, is only a {\em sufficient} and not a {\em necessary} condition for seeing such a deviation.

\robs This point was argued in the introduction of Ref.~\cite{mazurek21}.  Probing new regimes, which we termed the ``terra-nova'' strategy, is just one way to discover new physics; probing the precision frontier is another.\blk
Our experiment provides an opportunity to see deviations from operational quantum theory even if these are small  because there is always some level of precision at which a small deviation can be detected.  The fact that our experiment did not provide evidence in favour of such deviations implies that either (i) they are not there, or (ii) they are there, but can only be detected in an experiment that achieves higher precision.  




Finally, we would like to take this opportunity to correct something that HH state regarding the connection between overfitting and the notion of fine-tuning:
 \begin{quote} The only scientifically relevant notion of finetuning is overfitting. All other notions of finetuning are properties that a model may or may not have, but which do not a priori imply the model has a problem.
 \end{quote}
In the category of ``all other notions of fine-tuning, '' they include the failure of faithfulness described in Ref.~\cite{wood15}.
 HH here endorse two theses: (i) if a model fails to satisfy the assumption of faithfulness from Ref.~\cite{wood15},
 this does not imply that the model has a problem, and (ii) if a model overfits the data, then it {\em does} have a problem.
 However, 
  failing to satisfy the assumption of faithfulness generally {\em implies} overfitting, and so it is inconsistent to consider  the latter problematic and the former not.
This is explained  in Appendix B.1 of our article.  
Models that are structurally radical (such as quantum-extending versions of \textsc{cSD} models) can overfit the data relative to models that are structurally conservative (such as the \textsc{qCC}  model) because finite-run data always exhibits statistical fluctuations away from the no-signalling condition
and only in the \robs structurally radical models \blk
 can these fluctuations be mistaken for real features. \robs The fact that structurally radical models fail to be faithful is what opens up the possibility for them to interpret a violation of the no-signalling condition as a real feature and thereby overfit the data. \blk

\bibliographystyle{apsrev}
\bibliography{DaleyEtAlReply_BibFile}


\appendix

\section{Further comments on the relationship between model inclusion and overfitting}\label{app}

\subsection{An example of model inclusion without overfitting}

In Appendix C.3 of our article~\cite{daley22}, we described a dephased version of our experiment, that is, a Bell experiment where the entangled state is subject to dephasing and thus becomes a separable state, so that no Bell inequality violation is possible. \robs 
We then fit the data from this experiment to four distinct causal models and analyzed how they each performed relative to the train-and-test methodology.

We considered two causal models that are structurally conservative in the sense that they posit Bell's causal structure (a common cause acting on the pair of outcomes).  The distinction between the two is in the nature of the parameters that they add to this causal structure.   One is a model that is parametrically {\em quantum},
 denoted \textsc{qCC}).  The other is a model that is parametrically {\em classical},
  denoted \textsc{cCC}.  In other words, the first is an intrinsically quantum causal model, while the second is a classical causal model.   Because the dephased version of the experiment does not yield any Bell inequality violations, we expect that {\em both} the \textsc{qCC} and \textsc{cCC} models should be able to fit the data with a small training error.\footnote{This is the main difference between the dephased and nondephased versions of our experiment.  In the nondephased version, the source is entangled, leading to Bell inequality violations which the \textsc{cCC} model cannot fit well.} Furthermore, given that one can simulate a classical causal model using a quantum causal model, but not vice-versa, the \textsc{qCC} and \textsc{cCC} models stand in a relationship of model inclusion.  It is for this reason that the results reported in Appendix C.3 of our article are pertinent to the question of whether a more expressive model will generically be found to overfit the data relative to the other. 

In our analysis, we also considered two classical causal models that are structurally radical.  One model allows for a causal influence from a setting variable on one side of the experiment to the outcome variable on the other side. This model was termed {\em 
 structurally cause-effect} and was denoted \textsc{cCE}$_0$.  The other model allowed for the common cause to influence a setting variable in addition to influencing both outcome variables.  It was termed {\em 
  structurally superdeterministic} and was denoted \textsc{cSD}$_0$.


It turns out that there is a particular overfitting pitfall (i.e., a vulnerability to mistaking statistical fluctuations for real features) that arises {\em only } in the pair of structurally radical models. 

Note, first of all, that any finite-run sample of training data from the dephased version of the experiment will generally yield relative frequencies that exhibit statistical fluctuations {\em away from} the exact satisfaction of the no-signalling condition. (The dephased version of the experiment is like the original experiment in this regard.)
Both of the structurally radical causal models can mistake these fluctuations for real features because they both allow for choices of parameter values that imply a violation of the no-signalling condition. Consequently, the best-fit values of the parameters for these models (obtained from the training data) can lead to a prediction of a small violation of the no-signalling condition in the test data.  
The structurally conservative causal models, on the other hand, cannot mistake these fluctuations for real features because all choices of parameter values in these models lead to correlations that strictly satisfy the no-signalling condition.  As such, the best-fit values of the parameters for these models (obtained from the training data) always predict no violation of the no-signalling condition in the test data.  The relative frequencies computed from the test data will also exhibit statistical fluctuations away from the no-signalling condition, but these fluctuations are unlikely to be of the same type as those exhibited in the training data.  As such, one expects the structurally radical models to have a higher test error than the structurally conservative models. 

This expectation was born out by the analysis presented in Appendix C.3 of our article~\cite{daley22}, summarized in Fig. 6.   Specifically, we found that both of the structurally radical  models, \textsc{cCE}$_0$ and \textsc{cSD}$_0$, had a larger test error and a smaller training error than the structurally conservative models, \textsc{cCC} and \textsc{qCC}, hence the former two exhibited overfitting relative to the latter two. 

But for the question of the relationship between model inclusion and overfitting, it is a different comparison among the models that we wish to focus on here, namely, the comparison between the \textsc{cCC} and \textsc{qCC} models.  This is because the \textsc{qCC} model is strictly more expressive than than the 
\textsc{cCC} model, and yet does not overfit the data relative to the latter.

More precisely, under the condition that the cardinality of the latent variable acting as common cause in \textsc{cCC} is equal to the dimension of the latent quantum system acting as common cause in \textsc{qCC}, all data tables realizable by the \textsc{cCC} causal model can also be realized by the \textsc{qCC} causal model.  The results depicted in Fig. 6 of the appendix of our article were for a 2-bit common cause in the case of \textsc{cCC} and a 2-qubit common cause in the case of \textsc{qCC}.  Consequently, the \textsc{qCC} and \textsc{cCC} models being compared in this figure stand in a relationship of model inclusion.

Recall from the discussion above that {\em neither} of these models are able to mistake statistical fluctuations away from the no-signalling condition for real features, i.e., neither can exhibit the type of overfitting that the \textsc{cCE}$_0$ and \textsc{cSD}$_0$ models are vulnerable to.  Furthermore, we do not see 
  any other avenues for mistaking statistical fluctuations for real features in the dephased experiment that we realized, in particular, none to which a \textsc{qCC} model would be susceptible but not a  \textsc{cCC} model.\footnote{See, however, comments below regarding such a possibility for an experiment different from the one we conducted.}  Consequently, in the analysis of the experimental data, we do not expect the \textsc{qCC} model to overfit the data relative to the \textsc{cCC} model in spite of the model inclusion relation holding between them. 

This expectation was also born out by the data analysis, which was summarized in our Fig.~6: the  \textsc{qCC} model had training and test errors that are essentially the {\em same} as those of the \textsc{cCC} model. 
This, therefore, constitutes an example of model inclusion without overfitting.

The incorrect presumption made by HH is that {\em whenever} a model $M$ stands in a relationship of strict inclusion to a  model  $M'$ in the sense that it can reproduce predictions of $M'$ and more besides, then $M$ will overfit the data relative to $M'$.
But in fact the extra expressive power is sometimes innocuous, as is exemplified by the comparison of \textsc{qCC}  and \textsc{cCC}  for our dephased experiment.
It is innocuous when it does not lead to statistical fluctuations being mistaken for real features. The \textsc{qCC} model used in our data analysis certainly has additional expressive power relative to the \textsc{cCC} model we used, but our results demonstrate that, for the case of our experiment, it is of the innocuous variety.  
 
 In summary, HH's notion that model inclusion {\em necessarily} implies overfitting  is not right.  
This undermines their argument against our analysis technique.

\subsection{A novel opportunity for overfitting in a model inclusion setting}

Nonetheless, it is worth considering the question that this raises: 
what sorts of circumstances {\em could} lead to the situation that a more expressive model overfits the data relative to a less expressive model?  

To answer this question, we return to the example of a dephased version of a Bell experiment and describe novel circumstances (distinct from the one realized in the experiment we conducted) in which one {\em could} find that the \textsc{qCC} model overfits the data relative to the  \textsc{cCC} model using the data analysis technique we used for our experiment. Recall that both models satisfy the no-signalling condition for all parameter values and that the only difference between them is that the \textsc{qCC} model can violate Bell inequalities while the \textsc{cCC} model cannot.  Now suppose that one does a dephased version of the experiment wherein one targets a separable state and measurements that precisely {\em saturate} the bound in a Bell inequality. 
This ideal is never quite achieved by the separable state and measurements that one {\em actually} realizes in the experiment, because of unavoidable noise and imprecision. 
Suppose the noise and precision are such one expects that the correlations achieved fall short of saturating the Bell bound by $\epsilon$.  Suppose further that the amount of data collected in the experiment is 
such that the variance of the statistical fluctuations is expected to be larger than $\epsilon$.  In this case, there could be  a statistical fluctuation that leads to a violation of the Bell inequality for the finite-run relative frequencies of the training data.  Such a violation could be mistaken for a real feature by the  \textsc{qCC} model, such that the best-fit set of parameter values for this model might describe a small amount of entanglement in the prepared state rather than one that is separable. 
In its predictions about the test data, such a best-fit model would then assign a larger probability to a particular type of the Bell inequality violation.  On the other hand, because the  \textsc{cCC} model cannot violate a Bell inequality for any choice of parameter values, it cannot mistake such a fluctuation for a real feature and hence will not predict any such violation in the test data.  The conclusion is that  the \textsc{qCC} model could be found to have worse predictive power than the  \textsc{cCC} model in these circumstances.

It is worth contrasting this example of decreased predictive power to the example that was realized in our experiment. 
For this purpose, consider the comparison of two causal models that stand in a relationship of model inclusion, where one is structurally radical and the other is structurally conservative, such as the \textsc{cSD}$_0$ and the \textsc{cCC} models.
In this case, some of the extra expressive power of the former compared to the latter comes in the form of the possibility of a violation of an {\em equality} constraint, namely, the no-signalling condition.  
This sort of difference does not arise between the \textsc{qCC} and \textsc{cCC} models, because these satisfy the same equality constraints. 
Rather, the difference in the set of correlations that can be realized in \textsc{qCC} and \textsc{cCC} models is only relative to {\em inequality} constraints.   Specifically, the Bell inequalities are satisfied by the \textsc{cCC} model but can be violated by the \textsc{qCC} model.  

For generic causal structures (in either classical or quantum causal models), one can have  both equality and inequality constraints~\cite{wolfe2019inflation}. 
 What the discussion above demonstrates is that differences in the {\em inequality} constraints satisfied by a pair of models $M$ and $M'$ that stand in a relationship of model inclusion
 may constitute another avenue by which a statistical fluctuation can be mistaken for a real feature in $M$ but not in $M'$.  Hence, it is another avenue by which $M$ might lead to overfitting of the data relative to $M'$.  Note, however, that this is only expected to occur when the data is exceptional, lying within statistical error of the boundary between the set of observational data realizable by $M'$ and the set of observational data that is only realizable by $M$. 

%
%

The dephased experiment that we performed had a sufficiently small statistical error (relative to the imprecision in the state preparation and measurements) that fluctuations leading to a violation of the Bell bound were very unlikely.
 We consequently did not expect to see any evidence of the \textsc{qCC} model overfitting relative to the \textsc{cCC} model, and indeed we saw none. \blk

Nonetheless, it is worth addressing the question of what we ought to conclude from the possibility of an alternative experimental circumstance that would allow for such overfitting.  Do such considerations constitute a {\em reductio ad absurdum} argument against the train-and-test methodology for model selection? 

They do not.  This is because it is always possible to include more information about the experiment when finding the best-fit model, and doing so generally prevents the more expressive model from falling into the trap of mistaking statistical fluctuations for real features.




Consider the case of the dephased version of our experiment.  In our data analysis, we conceptualized the initial preparation procedure for the bipartite state, call it $P$, as a black box even though it was implemented as the composition of two procedures:
 (i) the  procedure that prepared an entangled state of the bipartite system, which was achieved using parameteric downconversion,  call it $P_0$, and (ii) the transformation on the bipartite system that realized the dephasing, which was achieved by implementing a uniform mixture of 
identity and Pauli $X$ operations, call it $T$.  Our fitting procedure looked for the bipartite quantum state that provided the best fit to the procedure $P$, the black-box conceptualization of the preparation.  However, one could opt for a fitting procedure that models the two components of the preparation procedure separately, that is, a fit to $P_0$ and a fit to $T$.  One could furthermore appeal to additional experimental data in achieving this fit.  For instance, \robs by preparing each of a tomographically complete set of states on the pair of systems (rather than confining the procedure $P_0$ to the preparation of a single entangled state), \blk  
one could achieve a tomographic characterization of the dephasing operation itself.  Doing so would yield experimental data that provided strong evidence 
 in favour of modelling $T$ as an entanglement-breaking channel, such that the bipartite state representing $P$, i.e., the composition of \robs the entangled state $P_0$ and the channel $T$,\blk  would in turn be 
  modelled by a separable state rather than one that is entangled. 



\robs In this revised version of the dephased experiment, the \textsc{qCC} model would be more likely to interpret a slight violation of Bell inequalities in the finite-run relative frequencies as a statistical fluctuation consistent with a state that is separable rather than a real feature necessitating a state that is entangled. \blk  In this way, such fluctuations would not be mistaken for real features and consequently the \textsc{qCC} model would not be found to have worse predictive power than the corresponding \textsc{cCC} model.  

This exemplifies how, in general, one avoids the {\em reductio} argument against the train-and-test methodology for comparing models that stand in a relationship of model inclusion to one another: 
by incorporating additional details of the experimental set-up, the more expressive model 
 can be prevented from mistaking statistical fluctuations for real features.


In principle, one could take this approach towards tests of theories that make predictions distinct from operational quantum theory.  

Consider, for example, Valentini's quantum nonequilibrium variant of Bohmian mechanics~\cite{valentini1991signal1,valentini1991signal2}.
 \robs It predicts the possibility of violations of the no-signalling condition.  
 Imagine an experiment has been done that sees such a violation in the finite-run relative frequencies, but where it is unclear whether it should be considered to be a real feature, hence evidence in favour of Valentini's hypothesis, or merely a statistical fluctuation. \blk   Including certain details of the realized experimental procedure may constrain the fit in a way that reduces the chances that the model mistakes statistical fluctuations for real features. For instance, \robs suppose the experiment is not probing the sort of exotic matter that Valentini argues is most likely to exhibit deviations from quantum equilibrium~\cite{valentini2004extreme}.  In this case, taking into account additional details of the experimental set-up might be sufficient to render unlikely any deviation from quantum equilibrium in the fit, such that the best-fit model interprets \blk
a violation of no-signalling in the finite-run data as a statistical fluctuation rather than a real feature.  

As another example, it has been proposed that we might some day find experimental evidence in favour of an alternative to quantum theory that allows for violations of Bell inequalities that are {\em larger} than the maximal violations realizable in quantum theory, i.e., one that allows for violations of the Tsirelson bound~\cite{navascues2015almost}.  The framework of causal models can be adapted to these alternative theories (see the discussion of GPT causal models in Ref.~\cite{wolfe2019inflation} for instance), and consequently one could apply the train-and-test methodology described here to try and adjudicate between two causal models that have the Bell causal structure: the quantum model and a post-quantum alternative to it. Suppose that the experiment was such that the quantum predictions are within statistical error of the Tsirelson bound. Assuming quantum theory is the correct model, there could nonetheless be a statistical fluctuation such that the relative frequencies computed from the finite-run training data violated the Tsirelson bound.  In this case, the post-quantum causal model, which allows for choices of parameter values that violate the Tsirelson bound, might mistake such a fluctuation for a real feature, and predict that the test data will exhibit a similar violation of the Tsirelson bound.  Meanwhile, the quantum causal model satisfies the Tsirelson bound for all choices of parameters and hence necessarily predicts that the test data will satisfy the Tsirelson bound.  In this way, the post-quantum model might be found to overfit the data relative to the quantum model. But if a proponent of a Tsirelson-violating alternative to quantum theory had a hypothesis about the particular experimental circumstances in which the violation was likely to occur, and the experiment in question was not of the right type, that additional information might be incorporated into the prior over the parameter values for the post-quantum model within the fitting procedure and it might ensure that the best-fit parameters obtained from the training data did not fall prey to the overfitting trap.  In this way, such a proponent could show that the particular experiment in question did not provide evidence against their hypothesis. 

To summarize, a given theory provides different parametric models of an experiment---in particular, different prior probability distributions over the space of parameter values---depending on the amount of detail in the experimental set-up one takes into account in the modelling procedure.  Consequently, even if a theory is diagnosed as overfitting the data  when the experiment is treated at one level of detail, this diagnosis might be overturned when additional information about the experimental procedure is taken into account. 





\blk

\end{document}